\documentclass[intlimits,twoside,a4paper]{article}

\usepackage[cp1251]{inputenc}
\usepackage[eqsecnum]{cmpj3}
\DeclareSymbolFont{yhlargesymbols}{OMX}{yhex}{m}{n}
\DeclareMathAccent{\wideparen}{\mathord}{yhlargesymbols}{"F3}
\issue{2022}{25}{4}{43707}
\doinumber{10.5488/CMP.25.43707}

\title[Slow and fast relaxation times of quantum lattice model with local multi-well potentials]
{Slow and fast relaxation times of quantum lattice model with local multi-well potentials: phenomenological dynamics for Sn$_{2}$P$_{2}$S$_{6}$ ferroelectric crystals}

\author[R. Erdem, S. \"{O}z\"{u}m, N. G\"{u}\c{c}l\"{u}]{R. Erdem\orcid{0000-0001-7370-2263}\refaddr{label1}\thanks{Corresponding author: \email{rerdem@akdeniz.edu.tr}.},
	S. \"{O}z\"{u}m\orcid{0000-0003-2123-5856}\refaddr{label2}, N. G\"{u}\c{c}l\"{u}\orcid{0000-0002-0069-0238}\refaddr{label3}}
\addresses{
	\addr{label1} Department of Physics, Akdeniz University, 07058, Antalya, T\"{u}rkiye
	\addr{label2} Alaca Avni \c{C}elik Vocational School, Hitit University, 19600, \c{C}orum, T\"{u}rkiye
	\addr{label3} Department of Physics Education, Necmettin Erbakan University, 42090, Konya, T\"{u}rkiye
}

\date{Received June 29, 2022, in final form October 16, 2022}
\begin{document}
	
	\maketitle

		\begin{abstract}
			As a continuation of the previously published work [Velychko O. V., Stasyuk I. V., Phase Transitions, 2019, \textbf{92}, 420], a phenomenological framework for the relaxation dynamics of quantum lattice model with multi-well potentials is given in the case of deformed Sn$_{2}$P$_{2}$S$_{6}$  ferroelectric lattice. The framework is based on the combination of statistical equilibrium theory and irreversible thermodynamics. In order to study these dynamics in a connected way we assume that the dipole ordering or polarization ($\eta$) and volume deformation ($u$) can be treated as fluxes and forces in the sense of Onsager theory. From the linear relations between the forces and fluxes, the rate equations are derived and characterized by two relaxation times ($\tau_{S}, \tau_{F}$) which describe the irreversible process near the equilibrium states. The behaviors of $\tau_{S}$  and $\tau_{F}$  in the vicinity of ferroelectric phase transitions are studied.

			\printkeywords
		\end{abstract}

		\section{Indroduction}
		\indent Sn$_{2}$P$_{2}$S$_{6}$ (SPS) ferroelectrics (FEs) is one of the prospective materials which presents spontaneous polarization determined by anharmonic potentials for the order parameter fluctuations. Intensive experimental and theoretical efforts have been devoted to the study of SPS crystals \cite{1,2,3,4,5,6,7,8,9,10,11,12,13,14,15,16}. Among the theories, quantum lattice model (QLM) is of special interest for the SPS investigations \cite{17,18,19,20,21,22,23,24,25}. It is a lattice system with three-well local potentials related to Blume-Emery-Griffiths model \cite{26}. Based on the QLM, the thermodynamics of the deformed SPS crystals has recently been studied in the presence of external pressure by Velychko and Stasyuk \cite{24}. For the monoclinic SPS-FEs, they  obtained a set of mean-field self-consistent equations which describe stable states at a given mechanical stress and determined self-consistency parameters using the experimental data. From the numerical solutions, the first- and second-order phase transitions between the ferroelectric (F) and paraelectric (P) phases were obtained as well as the tricritical point (TCP). 
		
		Although a few investigations are devoted to the statics of the QLM-based SPS crystals up till now, there has been no study to observe the relaxation dynamics of the system. In this work, we have made use of a simple phenomenological way to define the slow and fast relaxation times near the ferroelectric phase transitions in the SPS crystals. Such a study of relaxation dynamics has already been presented to observe the time dependent behaviours near the phase transitions and critical phenomena in various systems~\cite{27,28,29,30,31,32,33,34,35}. 
		
		In this study, giving a brief description of the QLM with local multi-well potentials and its phase transitions for Sn$_{2}$P$_{2}$S$_{6}$  crystals under the MFA and being motivated by  \cite{31,32,33,34,35}, we have focused on the simple relaxation study of the system using phenomenological approach. Particularly, we observe the dynamic effects appearing in the regions of first- and second-order phase transitions as well as in the vicinity of the tricritical point. The large positive values of relaxation time near the F-P phase transitions  once again confirm the divergence singularity at the continuous phase transition observed in other cooperative phenomena.

		\section{The model description and static properties}
		The quantum lattice model with multi-well potentials is simply described by the Hamiltonian \cite{24} 
		
		\setcounter{equation}{0}
		\begin{equation}
			\widehat{H}  = \sum_{i} \widehat{H}_{i} + \widehat{H}_{1} + \widehat{H}_{2},
		\end{equation}
		
		\noindent where the first, second and third terms are known as the single-site, the interaction and the deformation energy parts, respectively. Since these terms are explicitly given in  \cite{24} and \cite{25}, we skip the details here and proceed with mentioning the properties of the system at equilibrium. These are determined self-consistently using Gibbs free energy calculations. Letting, $N, J, \eta, \nu, c_{0}, u, \theta, \varepsilon, D$ and  $\sigma_{S}$ be the number of lattice points, the effective field acting on dipoles, dipole ordering parameter, volume related with one formula unit, volume elastic constant, deformation (or relative volume change), reduced temperature, the renormalization of the energy gap due to the deformation, the constant of an electron-deformational interaction and mechanical stress, respectively, the mean-field Gibbs free energy per site is given by the following formula \cite{24}
		
		\begin{equation}
			f_{\text{MF}} = \frac{G_{\text{MF}}}{N} = \frac{1}{2}J\eta^2 + \frac{1}{2}\nu c_{0} u^2 - \theta \ln \left[1+2\exp\left(-\frac{\varepsilon+Du}{\theta}\right)\cosh\left(\frac{J\eta}{2\theta}\right)\right]- \nu u \sigma_{s}. 
			\label{eq2.2}
		\end{equation}
		
		\noindent Here, we define $J= \sum_{j} J_{ij}, \eta = \langle s_{i} \rangle$ ($s_{i}$ variable related to the local dipole moment, $\langle ... \rangle$ denotes the thermal expectation value), $\theta = kT$ ($k$ denotes the Boltzmann constant, $T$ means the absolute temperature), $\sigma_{S} = -p$ ($p$ is the hydrostatic pressure). The minimization of equation (\ref{eq2.2}) with respect to the variables $\eta$ and $u$ are treated as two independent variational parameters and using (\ref{eq2.2}) in the MFT give the following conditions
		
		\begin{equation}
			\frac{\partial f_{\text{MF}}}{\partial \eta} = 0, \quad \frac{\partial f_{\text{MF}}}{\partial u} = 0.
		\end{equation}
		
		\noindent For the system at equilibrium, the self-consistent equations are expressed in the form
		
		\begin{equation}
			\eta = \frac{\re^{-y}z}{1+2 \re^{-y}x}, 
		\end{equation}
		
		\begin{equation}
			\sigma = u c_{0} + \frac{D}{\nu} \left(\frac{2 \re^{-y}x}{1+2 \re^{-y}x}\right),
		\end{equation}
		
		\noindent where $x=\cosh\left(\frac{J\eta}{2\theta}\right)$, $y= \frac{\varepsilon+Du}{\theta}$, $z = \sinh\left(\frac{J\eta}{2\theta}\right)$. In order to have an insight into the phase transitions undergoing in the SPS-FEs on the monoclinic lattice structure, Velychko and Stasyuk \cite{24} firstly solved the above equations using the parameter values of $J=0.14$ eV, $\nu = 0.23 \times 10^{-21}$ cm$^3$, $c_{0} = 5 \times 10^{11}$~erg/cm$^{3}$, $\nu c_{0} = 71.8$ eV, $D = -1.1$ eV. They obtained the polarization $\eta$  and deformation $u$  as functions of the energy gap $\varepsilon$  and pressure $p$  at different temperature values $\theta$.  Recently, Erdem \cite{25} improved their results by adding $\eta$  vs. $\theta$  and $u$  vs. $\theta$  at a fixed pressure to represent the effect of temperature. Here, we summarize the basic results from these references for the convenience of our later discussions as follows: a finite jump of $\eta$  and $u$  at the first-order (discontinuous) phase transition from F phase to P phase accompanied by compression of the lattice occured. These jumps reduce as the TCP is approached. Above this point the phase transition becomes of the second-order (continuous), and thus the jumps of the parameters $\eta$  and $u$  vanish.

		\section{Theoretical framework}
		
		In order to study the relaxation dynamics of the above system we now assume that a small external stimulation is applied removing the system slightly from equilibrium. The Gibbs free energy produced in the irreversible process ($\Delta f_{\text{MF}}$) is calculated to observe how rapidly the system relaxes back to equilibrium. In this case, the Gibbs free energy near the equilibrium states will be
		
		\begin{equation}
			f_{\text{MF}}(T,\eta,u) = f^{(0)}_{\text{MF}}(T, \eta_{0}, u_{0}) + \Delta f_{\text{MF}},
		\end{equation}
		
		\noindent where $f^{(0)}_{\text{MF}}(T, \eta_{0}, u_{0})$ is the equilibrium free energy [found from  (\ref{eq2.2}) by setting $\eta = \eta_{0}, u = u_{0}$]. $\Delta f_{\text{MF}}$ is given by the terms in a Taylor series expansion of $f_{\text{MF}}$ with respect to the spontaneous equilibrium point $\eta = \eta_{0}, u = u_{0}$ as
		
		\begin{equation}
			\Delta f_{\text{MF}} = \frac{1}{2}\left[\phi_{\eta \eta} \left(\eta-\eta_{0}\right)^2 + 2\phi_{\eta u } \left(\eta - \eta_{0}\right)\left(u-u_{0}\right)+\phi_{u u}\left(u-u_{0} \right)^2 \right]. 
			\label{eq3.2}
		\end{equation}
		
		\noindent From  (\ref{eq3.2}), the expressions given below for $\phi_{\eta \eta}, \phi_{\eta u}, \phi_{u u}$ are explicit functions of the known equilibrium quantities $\eta_{0}, u_{0}$ and take the form
		
		\begin{equation}
			\phi_{\eta \eta} = \left(\frac{\partial^2 f_{\text{MF}}}{\partial \eta^2}\right)_{\text{eq}}, \quad  \phi_{\eta u} = \left(\frac{\partial^2 f_{\text{MF}}}{\partial \eta \partial u}\right)_{\text{eq}} = \phi_{u \eta} = \left(\frac{\partial^2 f_{\text{MF}}}{\partial u \partial \eta}\right)_{\text{eq}}, \quad \phi_{u u} = \left(\frac{\partial^2 f_{\text{MF}}}{\partial u^2}\right)_{\text{eq}}. \quad 
		\end{equation}
					
		\noindent Here, the subscribe `eq' means equilibrium. The generalized forces ($X_{\eta}, X_{u}$) can be described using the derivatives with respect to deviations ($\eta-\eta_{0}, u-u_{0}$), respectively:
		
		\begin{equation}
			X_{\eta} = -\frac{\partial (\Delta f_{\text{MF}})}{\partial (\eta-\eta_{0})} = -\phi_{\eta \eta} \left(\eta -\eta_{0} \right) - \phi_{\eta u} \left(u - u_{0} \right),
				\label{eq3.4}
		\end{equation}
		
		\begin{equation}
			X_{u} = -\frac{\partial (\Delta f_{\text{MF}})}{\partial (u-u_{0})} = -\phi_{u \eta} \left(\eta -\eta_{0} \right) - \phi_{u u} \left(u - u_{0} \right).
				\label{eq3.5}
		\end{equation}
		
		\noindent Based on the Onsager reciprocity theorem (ORT) \cite{36,37}, a linear relation between the currents ($\dot \eta, \dot u$) and forces ($X_{\eta}, X_{u}$) is introduced as follows in terms of a matrix of phenomenological rate coefficients (or Onsager constants):
		
		\begin{equation}
			\begin{aligned}
				&\begin{bmatrix}
					\dot{\eta}\\           
					\dot{u}
				\end{bmatrix}&
				=-
				&\begin{bmatrix}
					\gamma_{\eta}&\gamma\\           
					\gamma&\gamma_{u}
				\end{bmatrix}&
				&\begin{bmatrix}
					X_{\eta}\\           
					X_{u}
				\end{bmatrix},&
			\end{aligned}
		\end{equation}
		
		\noindent where the corresponding off-diagonal elements on the two sides of the main diagonal have the same signs (the matrix is symmetric since both $\eta$ and $u$  are even variables under time inversion), and the dot denotes a derivative with respect to time $t$.  The above matrix equation yields, upon using equations (\ref{eq3.4}) and (\ref{eq3.5}), the rate equations:
		
		\begin{equation}
			\dot \eta = \frac{\rd \eta}{\rd t} = -\Phi_{\eta \eta} \left(\eta -\eta_{0} \right) - \Phi_{\eta u} \left(u - u_{0} \right),
			\label{eq3.7}
		\end{equation}
		
		\begin{equation}
			\dot u = \frac{\rd u}{\rd t} = -\Phi_{u \eta} \left(\eta -\eta_{0} \right) - \Phi_{u u} \left(u - u_{0} \right).
			\label{eq3.8}
		\end{equation}
		\noindent The coefficients are defined by:
		\begin{equation}
			\Phi_{\eta \eta} = \gamma_{\eta} \phi_{\eta \eta} + \gamma \phi_{\eta u}, \quad \Phi_{\eta u} = \gamma_{\eta} \phi_{\eta u} + \gamma \phi_{u u}, \quad \Phi_{u \eta} = \gamma \phi_{\eta \eta} + \gamma_{u} \phi_{\eta u}, \quad \Phi_{u u} = \gamma_{u} \phi_{u u} + \gamma \phi_{\eta u}.
		\end{equation}
		\noindent Assuming a solution of the form $\exp(-t/\tau)$ for equations (\ref{eq3.7}) and (\ref{eq3.8}), one obtains the secular equation
		\begin{equation}
		\begin{aligned}
			\begin{vmatrix}
				\tau^{-1}-\Phi_{\eta\eta} & -  \Phi_{\eta u} \\ 
				-\Phi_{u \eta} & \tau^{-1} - \Phi_{uu}
			\end{vmatrix}
			=0,
		\end{aligned}
		\end{equation}

		\noindent which yields two inverse relaxation times \cite{38}:

		\begin{equation}
		\frac{1}{\tau_{S}} = \frac{1}{2} (\Phi_{\eta \eta} +\Phi_{u u})- \frac{1}{2}\big[(\Phi_{\eta \eta} -\Phi_{u u})^2+4 \Phi_{\eta u} \Phi_{u \eta}\big]^{1/2}, 
		\end{equation}		
		\begin{equation}
		\frac{1}{\tau_{F}} = \frac{1}{2} (\Phi_{\eta \eta} +\Phi_{u u})+ \frac{1}{2}\big[(\Phi_{\eta \eta} -\Phi_{u u})^2+4 \Phi_{\eta u} \Phi_{u \eta}\big]^{1/2}. 
		\end{equation}
 	
		\noindent Here, $\tau_{S}$  corresponds to a slower relaxation process while $\tau_{F}$ corresponds  to the faster one. They characterize the dipole ordering parameter and deformation (or relative volume change) relaxations, respectively, in the phenomenological approach using the ORT.

		\section{Numerical results}
		
			\begin{figure}[!b]
			\centering
			%\subfloat
			{\includegraphics[width=0.35\textwidth]{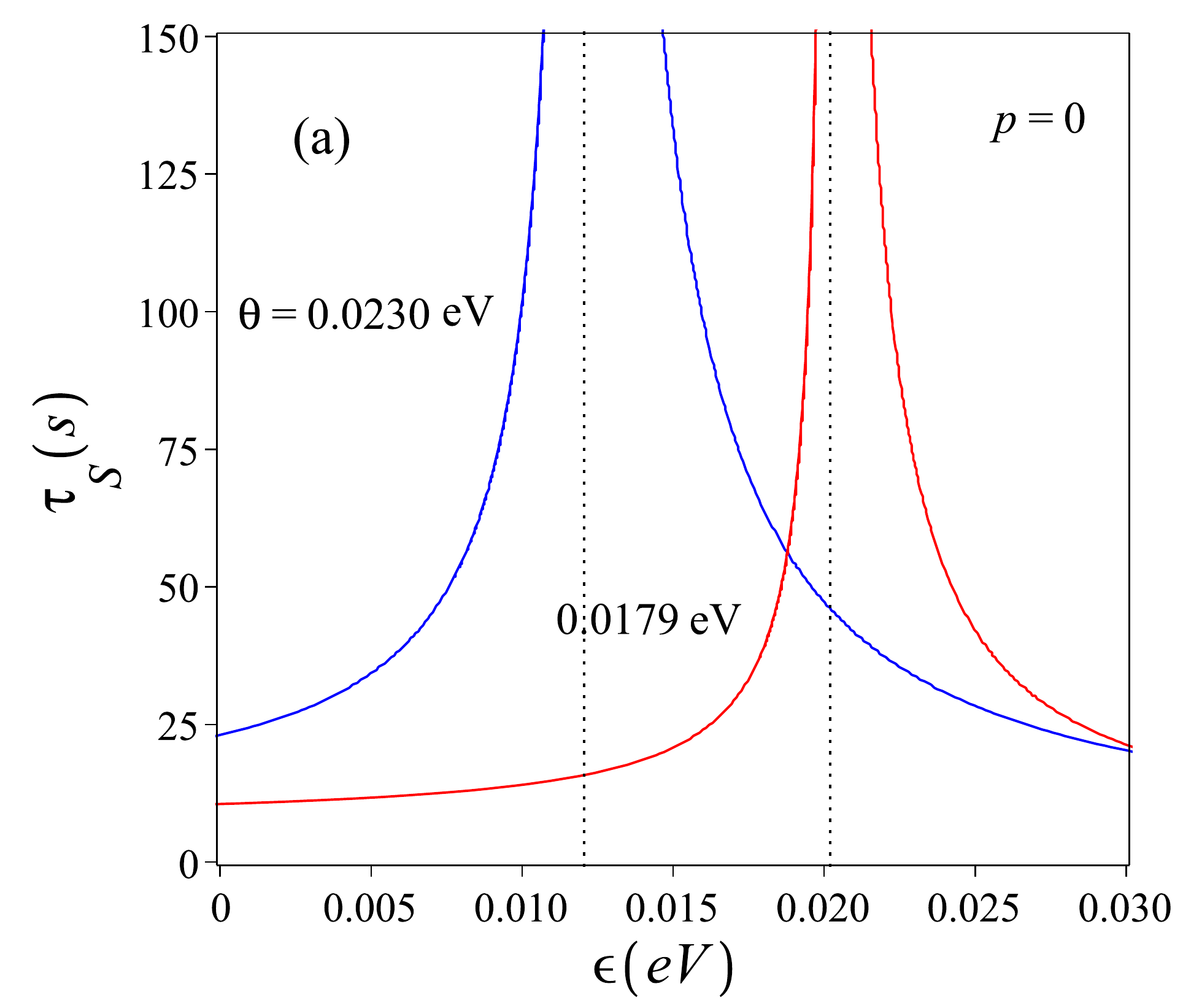}}
			%\subfloat
			{\includegraphics[width=0.35\textwidth]{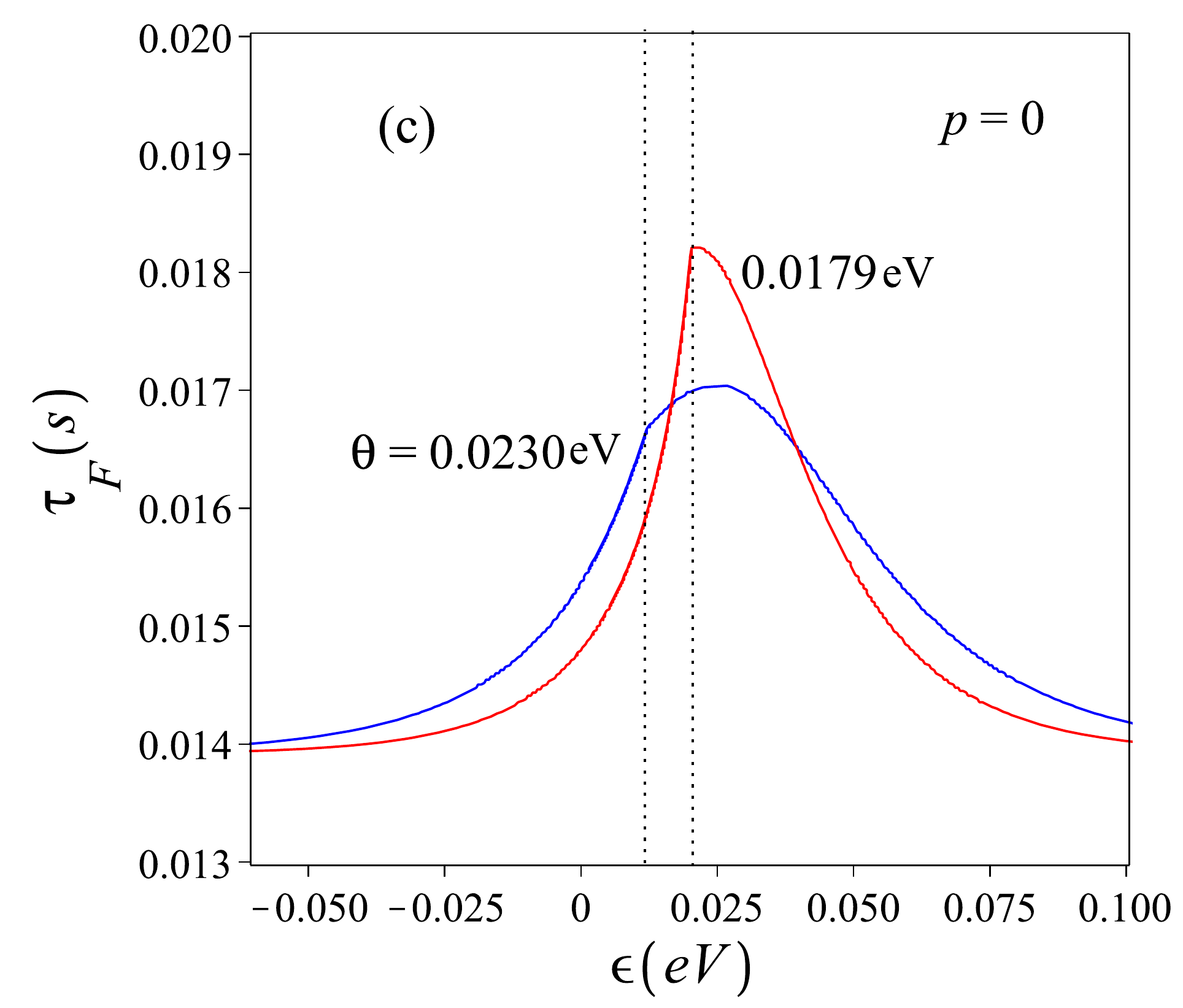}}\\
			%\subfloat
			{\includegraphics[width=0.35\textwidth]{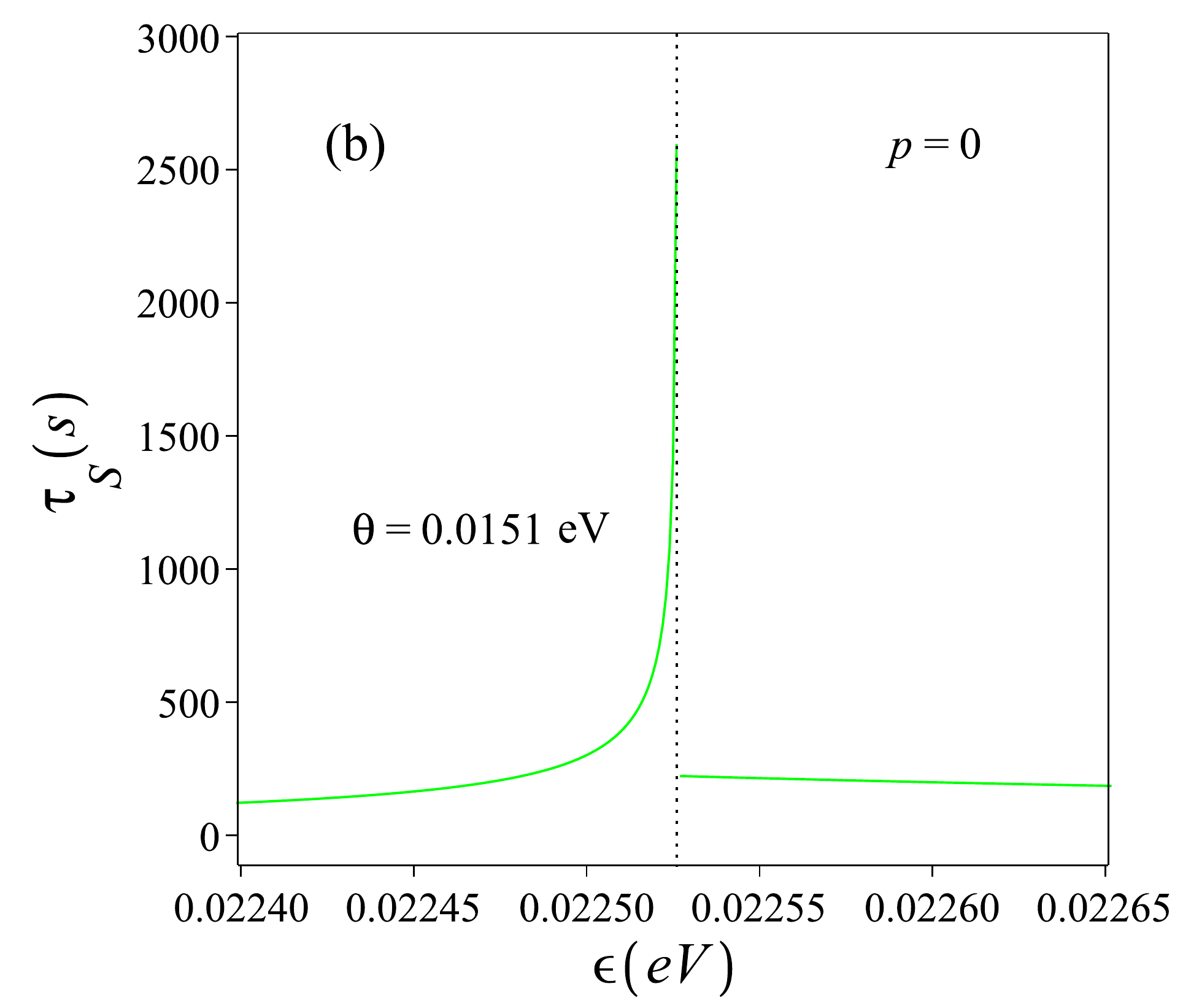}}
			%\subfloat
			{\includegraphics[width=0.35\textwidth]{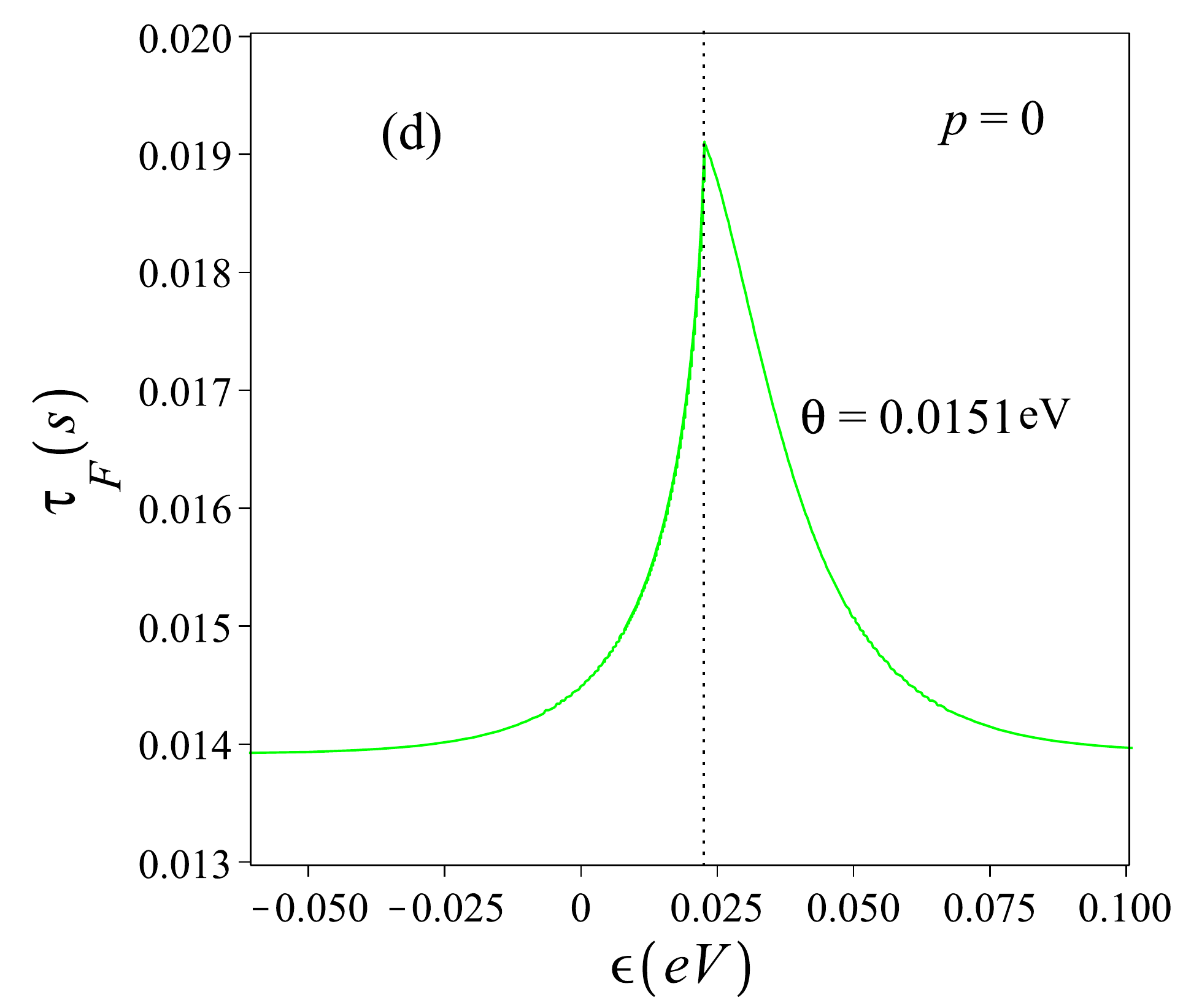}}
			\caption{(Colour online) (a), (b) Slow relaxation time $\tau_{S}$ and (c), (d) fast relaxation time $\tau_{F}$ vs. energy gap $\varepsilon$ at $\theta = 0.0230, 0.0179, 0.0151$~eV for $\gamma_{\eta}=\gamma_{u}=1$~eV$^{-1}$s$^{-1}$, $\gamma = 10^{-5}$~eV$^{-1}$s$^{-1}$.}
			\label{fig1}
		\end{figure}
		
		Firstly, the behavior of slow ($\tau_{S}$) and fast ($\tau_{F}$) relaxation times [in seconds ($s$) unit] as a function of the energy gap $\varepsilon$  (in electronvolts (eV) unit) is shown in figure~\ref{fig1} for several values of $\theta$ when there is no external pressure $p=0$. For Onsager rate coefficients, we choose $\gamma_{\eta}=\gamma_{u}=1$~eV$^{-1}$s$^{-1}$, $\gamma=10^{-5}$~eV$^{-1}$s$^{-1}$. The temperature values on the figures are for the second-order phase transition, the first-order phase transition and TCP. The vertical dotted lines refer to the phase transition values of $\varepsilon$, i.e., $\varepsilon_{C}$, $\varepsilon_{D}$ and $\varepsilon_{\text{TCP}}$. In this case, $\tau_{S}$  increases rapidly while rising (lowering) $\varepsilon$  value on left (right) side of $\varepsilon_{C}/\varepsilon_{\text{TCP}}$ and diverges to infinity, illustrated by the blue/red curves in figure~\ref{fig1}(a). On the contrary, it presents a large and abrupt drop across the discontinuous phase transition point [see figure~\ref{fig1}(b)]. Former singularity in $\tau_{S}$  is an expected property known as the signature of continuous phase transitions and hence well agrees with earlier investigations \cite{31,32}. However, the latter one is a novel property which is not found before. As for the fast relaxation time $\tau_{F}$, it also increases with $\varepsilon$ but remains scarcerly varied just around the continuous phase transition (tricritical point), as also shown via blue (red) curves in figure~\ref{fig1}(c). Hence, both critical and tricritical behaviours of $\tau_{F}$  are different from $\tau_{S}$, just a cusp singularity. Beyond $\varepsilon_{C}$, a maximum of $\tau_{F}$  is seen but it disappears at  $\varepsilon_{\text{TCP}}$. It should be stressed that cusps also occurred for $\tau_{F}$   during the discontinuous phase transition, indicated in figure~\ref{fig1}(d). These findings are also in agreement with previous relaxation studies \cite{32,35}.
				
			\begin{figure}[!t]
			\centering
			%			\subfloat
			{\includegraphics[width=0.35\textwidth]{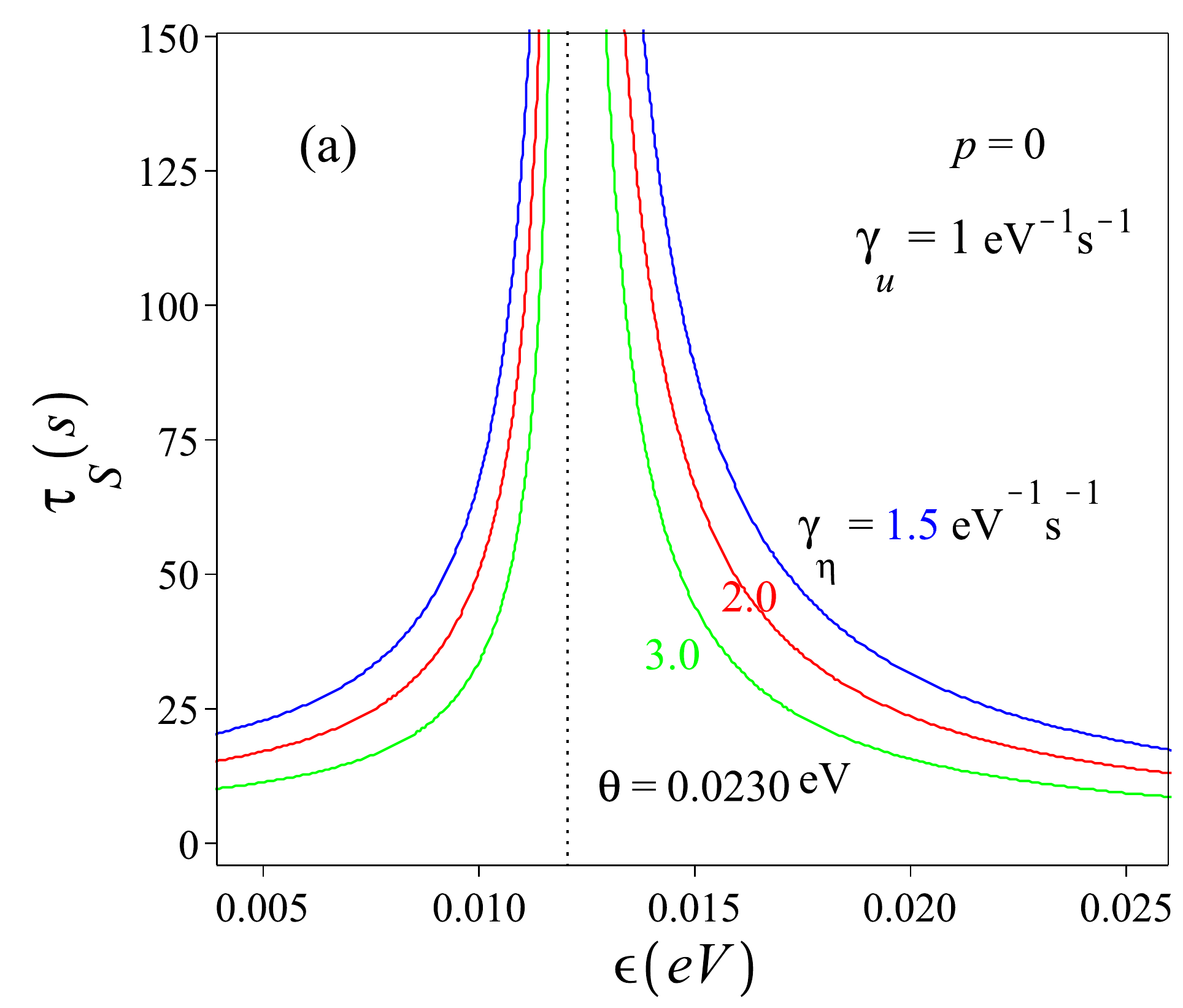}}
			%			\subfloat
			{\includegraphics[width=0.35\textwidth]{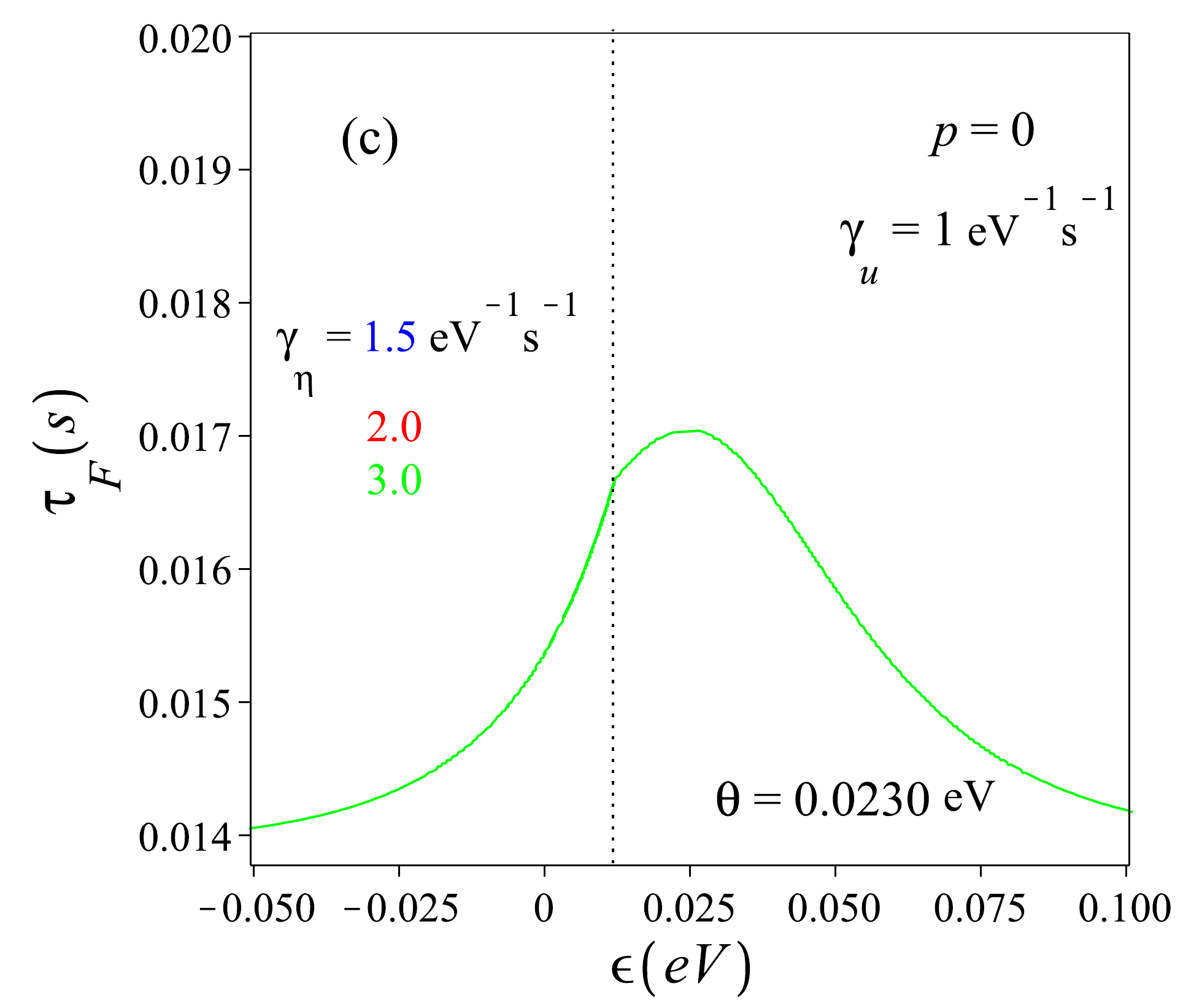}}\\
			%			\subfloat
			{\includegraphics[width=0.35\textwidth]{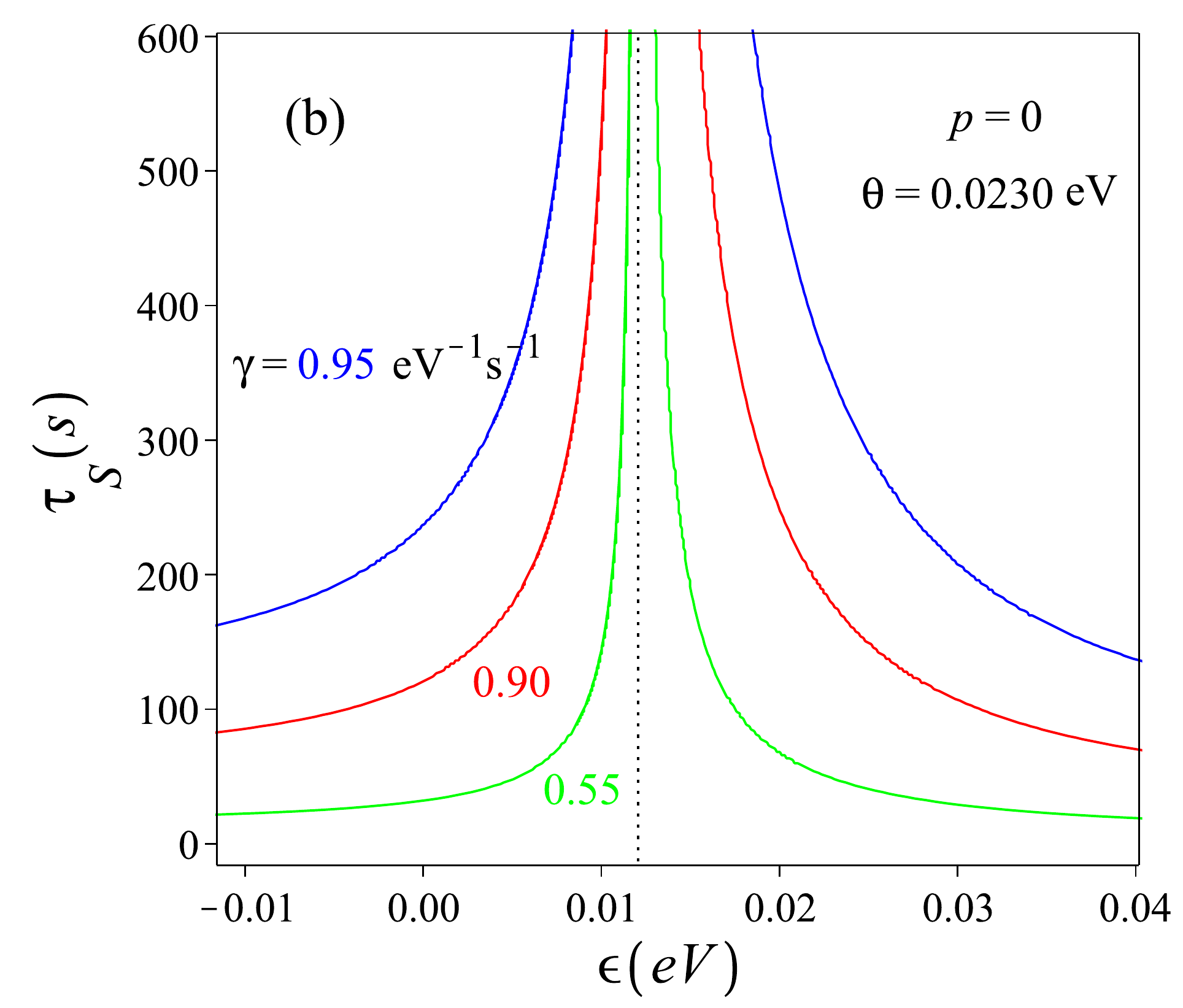}}
			%			\subfloat
			{\includegraphics[width=0.35\textwidth]{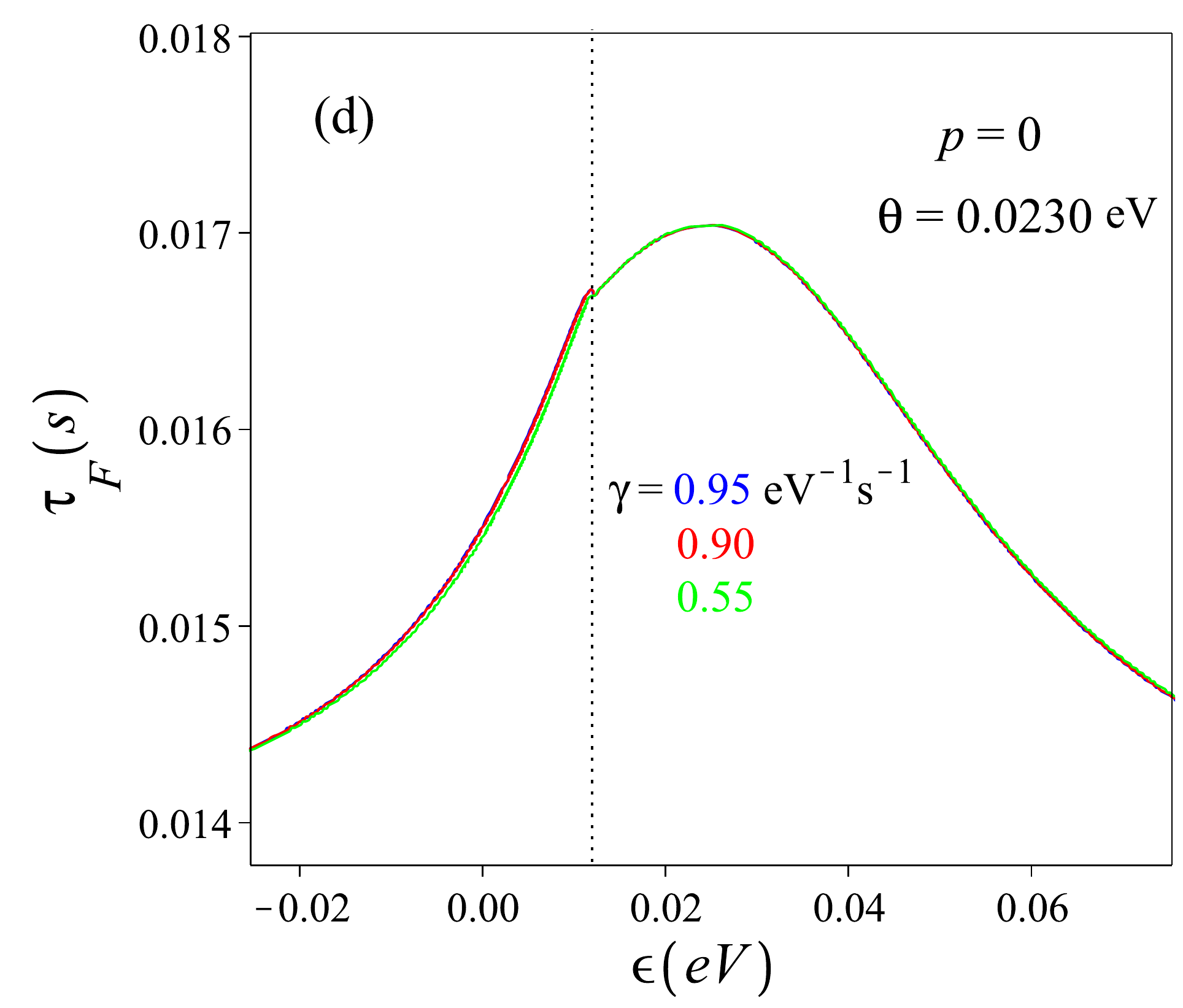}}
			\caption{(Colour online) (a), (b) Slow relaxation time $\tau_{S}$ and (c), (d) fast relaxation time $\tau_{F}$ vs. energy gap $\varepsilon$ in the absence and presence of the off-diagonal rate coefficient when $p=0, \theta = 0.0230$ eV.}
			\label{fig2}
		\end{figure}
		\begin{figure}[!t]
			\centering
			%	\subfloat
			{\includegraphics[width=0.35\textwidth]{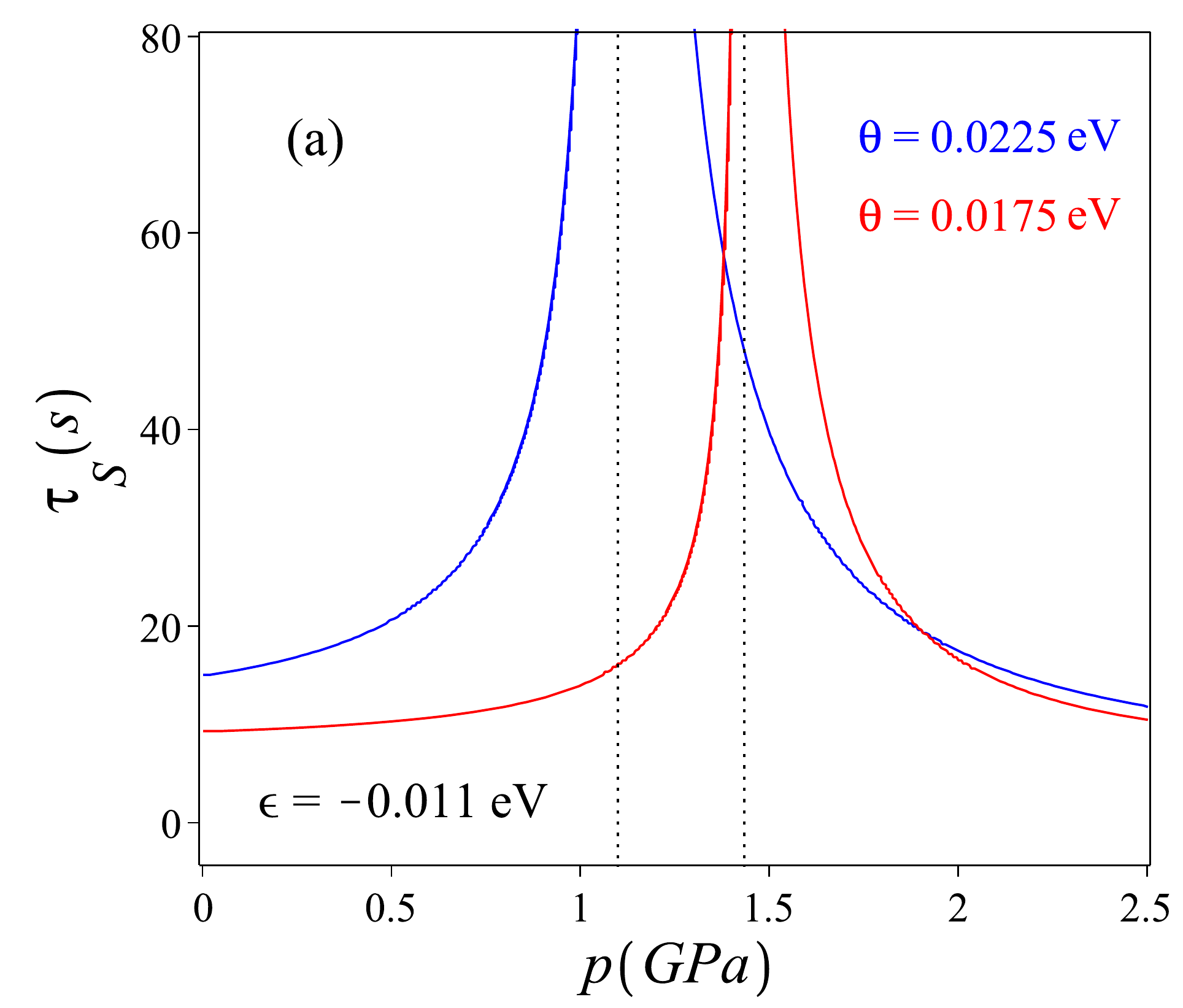}}
			%	\subfloat
			{\includegraphics[width=0.35\textwidth]{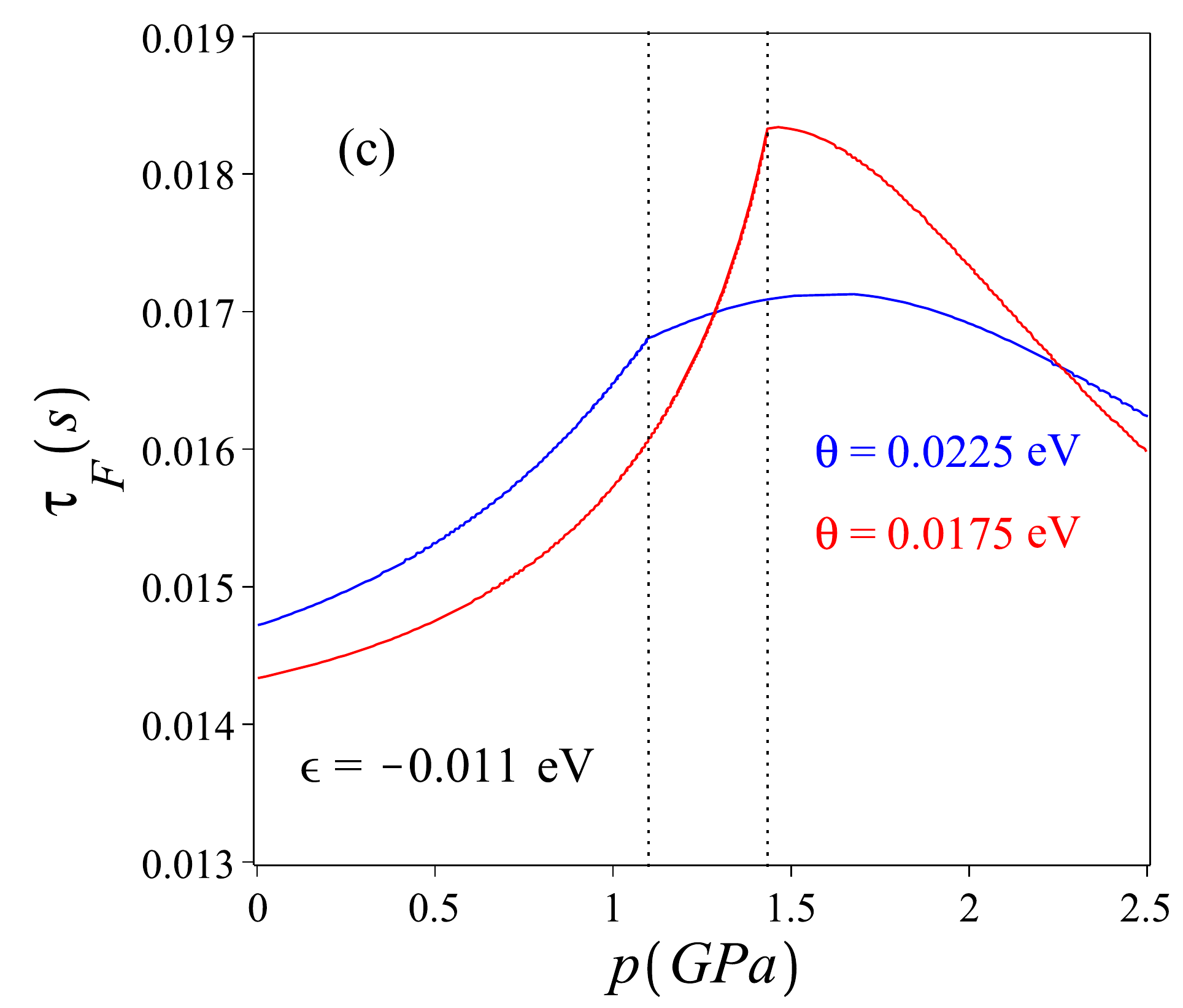}}\\
			%	\subfloat
			{\includegraphics[width=0.35\textwidth]{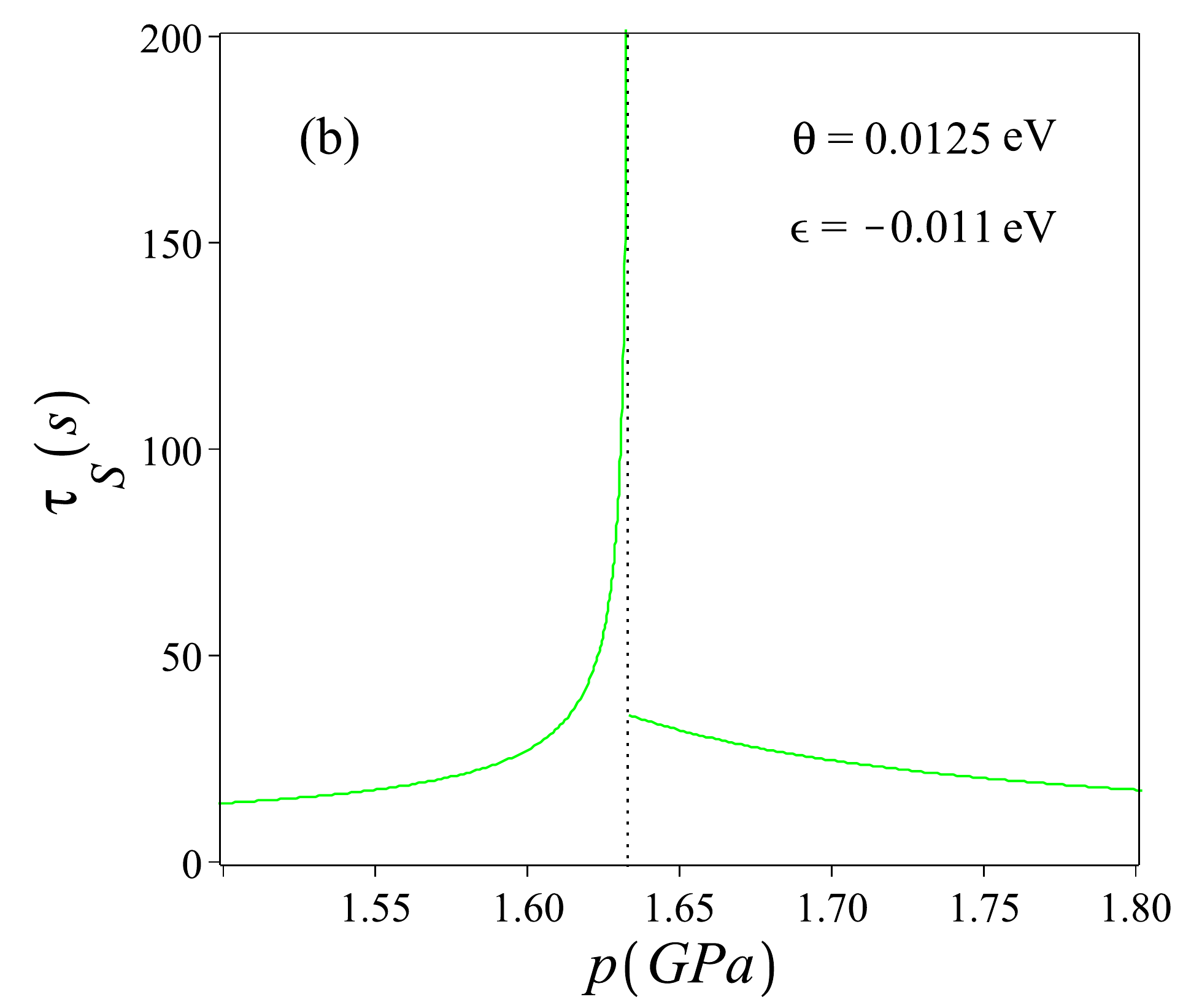}}
			%	\subfloat
			{\includegraphics[width=0.35\textwidth]{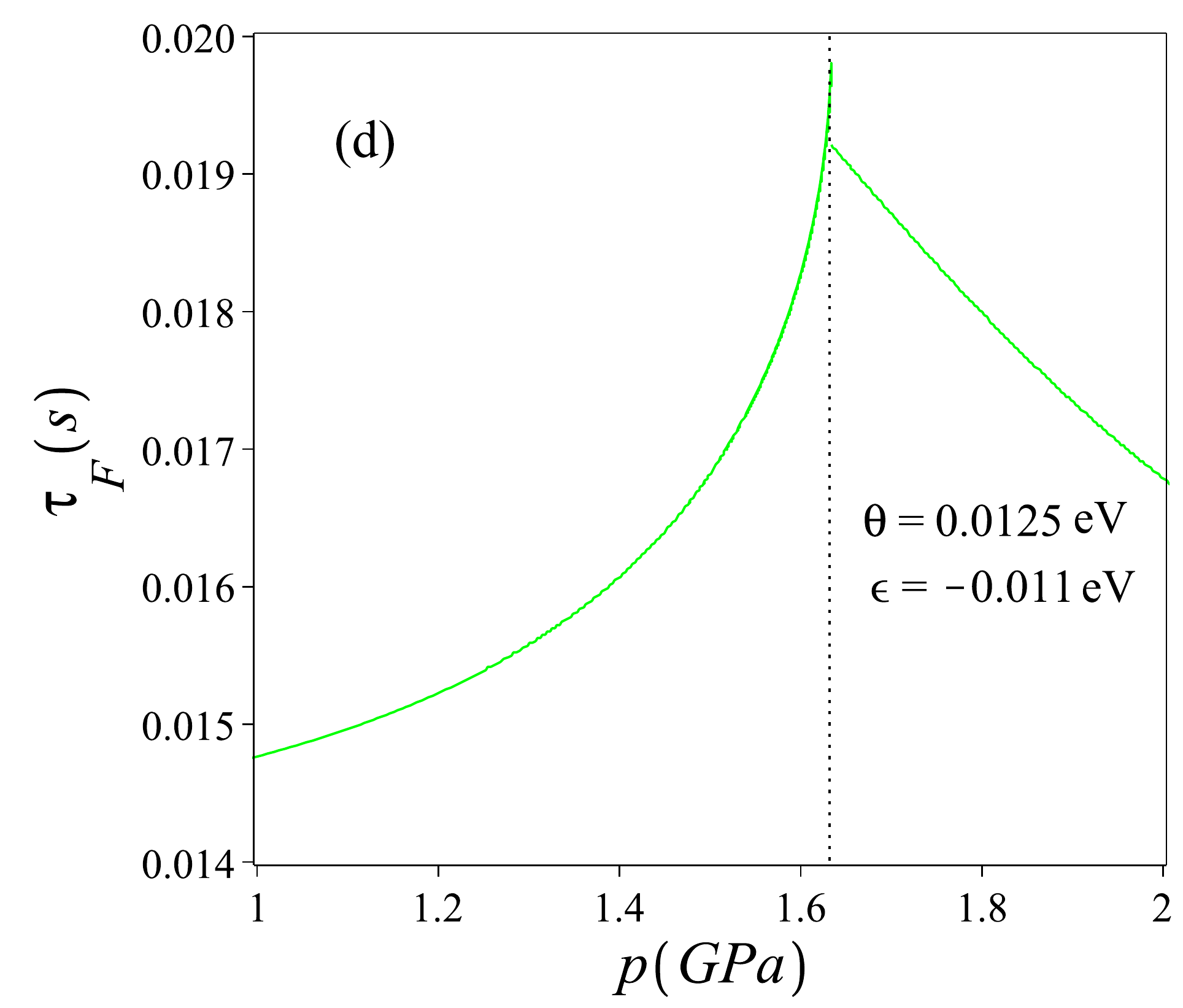}}
			\caption{(Colour online) (a), (b) Slow relaxation time $\tau_{S}$ and (c), (d) fast relaxation time $\tau_{F}$ vs. pressure $p$ at $\theta=0.0225, \; 0.0175, \; 0.0125 \; eV$ for $\gamma_{\eta}=\gamma_{u}=1$ eV$^{-1}$s$^{-1}$, $\gamma = 10^{-5}$ eV$^{-1}$s$^{-1}$.}
			\label{fig3}
		\end{figure}

		In order to show the effects of rate constants on the relaxation process, we illustrate in figure~\ref{fig2} the energy gap $\varepsilon$   variation of $\tau_{S}$  and $\tau_{F}$  in the absence and in the presence of the off-diagonal Onsager constant for $p=0$ and $\theta = 0.0230$ eV. As seen in  figure~\ref{fig2}(a), on both sides of $\varepsilon_{C}$, illustrated by a vertical dotted line, the divergence of $\tau_{S}$  gets pushed away from the critical energy gap value as we decrease the values of $\gamma_{\eta}$. This means that the rising of $\gamma_{\eta}$  leads to a speed up of the whole relaxation process. The opposite is valid in the case of off-diagonal Onsager coefficient $\gamma$ in figure\ref{fig2}(b). $\tau_{F}$  vs. $\varepsilon$ is shown for different values of $\gamma_{\eta}$ and $\gamma$  for $p=0$  and $\theta=0.0230$ eV in figures \ref{fig2}(c) and \ref{fig2}(d), respectively. It is evident that $\tau_{F}$ curves are independent of $\gamma_{\eta}$ and $\gamma$ values, shown in figures \ref{fig2}(c) and \ref{fig2}(d).

		Next, $\tau_{S}$  and $\tau_{F}$  vs. pressure $p$  [in gigapascal (GPa) unit] is displayed in figure~\ref{fig3} for three different temperatures using $\gamma_{\eta}=\gamma_{u}=1$~eV$^{-1}$s$^{-1}$, $\gamma=10^{-5}$ eV$^{-1}$s$^{-1}$. Figure\ref{fig3} also displays the pressure dependence of $\tau_{S}$  and $\tau_{F}$  where we adjust the parameters so as to obtain an agreement with figures~6(a), 6(b) in  \cite{25}. Our calculated dependence of ($\tau_{S}, \tau_{F}$) on $p$  demonstrates a good correspondence to previous $\varepsilon$  dependence of ($\tau_{S}, \tau_{F}$) in figure~\ref{fig1}. Our plots of the critical and tricritical cases are presented in figures~\ref{fig3}(a), \ref{fig3}(c) (blue and red curves) and the first-order case in figures~\ref{fig3}(b), \ref{fig3}(d) (green curves). In other words, the overall critical and tricritical behaviours of ($\tau_{S}, \tau_{F}$) do not change so much at the given $\varepsilon$  and $\theta$ values. However, the amount of the finite jump in $\tau_{S}$  at the first-order phase transition point seen in figure~\ref{fig3}(b) is very much smaller than that of figure~\ref{fig1}(b). It is important to mention that, because of the similar behavior of figure~\ref{fig2}, the pressure $p$  variation of $\tau_{S}$  and $\tau_{F}$  in the absence and in the presence of $\gamma$  for $p=0$  and $\theta=0.0230$ eV has not been drawn here.
		
		\begin{figure}[!t]
		\centering
		%			\subfloat
		{\includegraphics[width=0.35\textwidth]{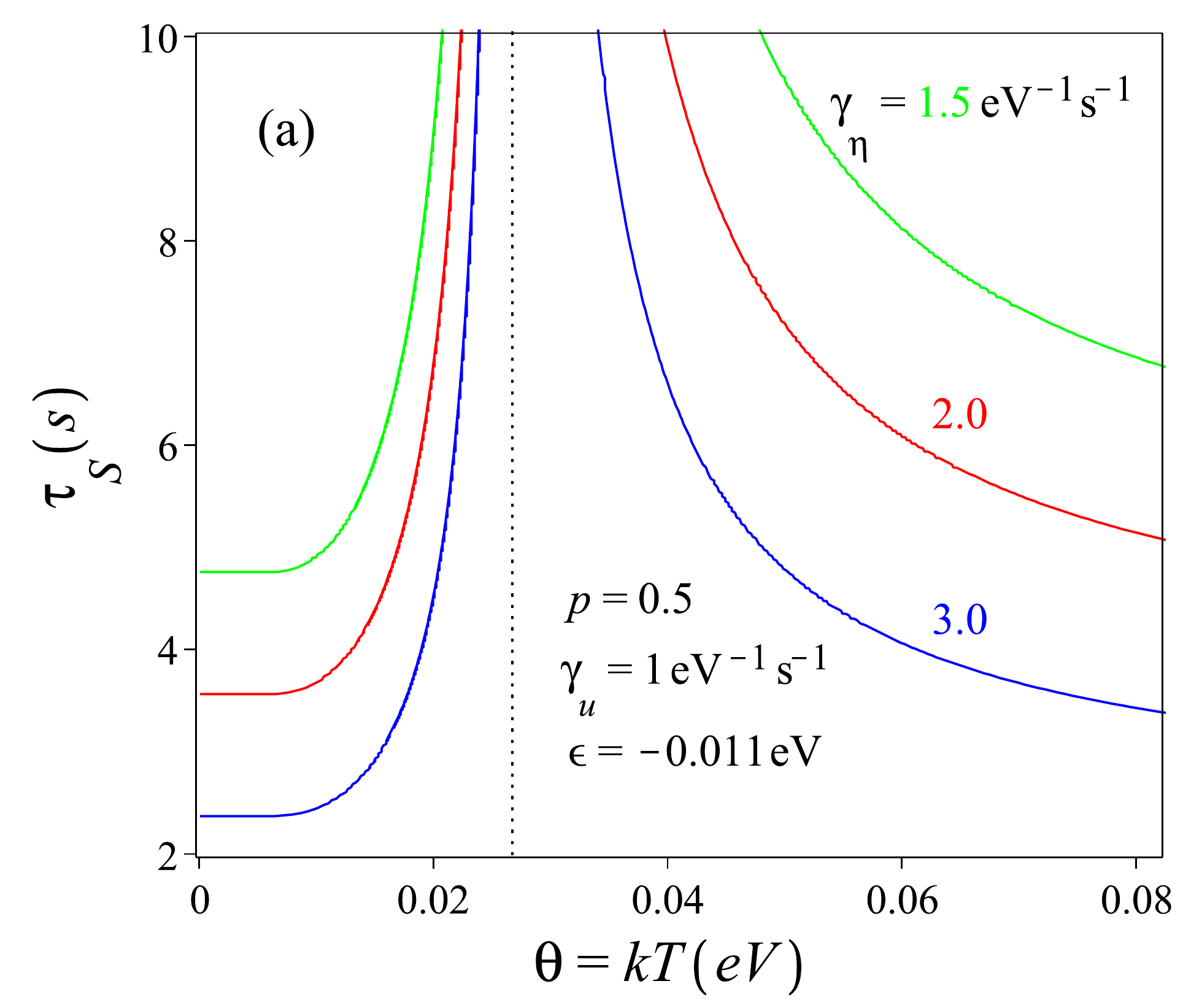}}
		%			\subfloat
		{\includegraphics[width=0.35\textwidth]{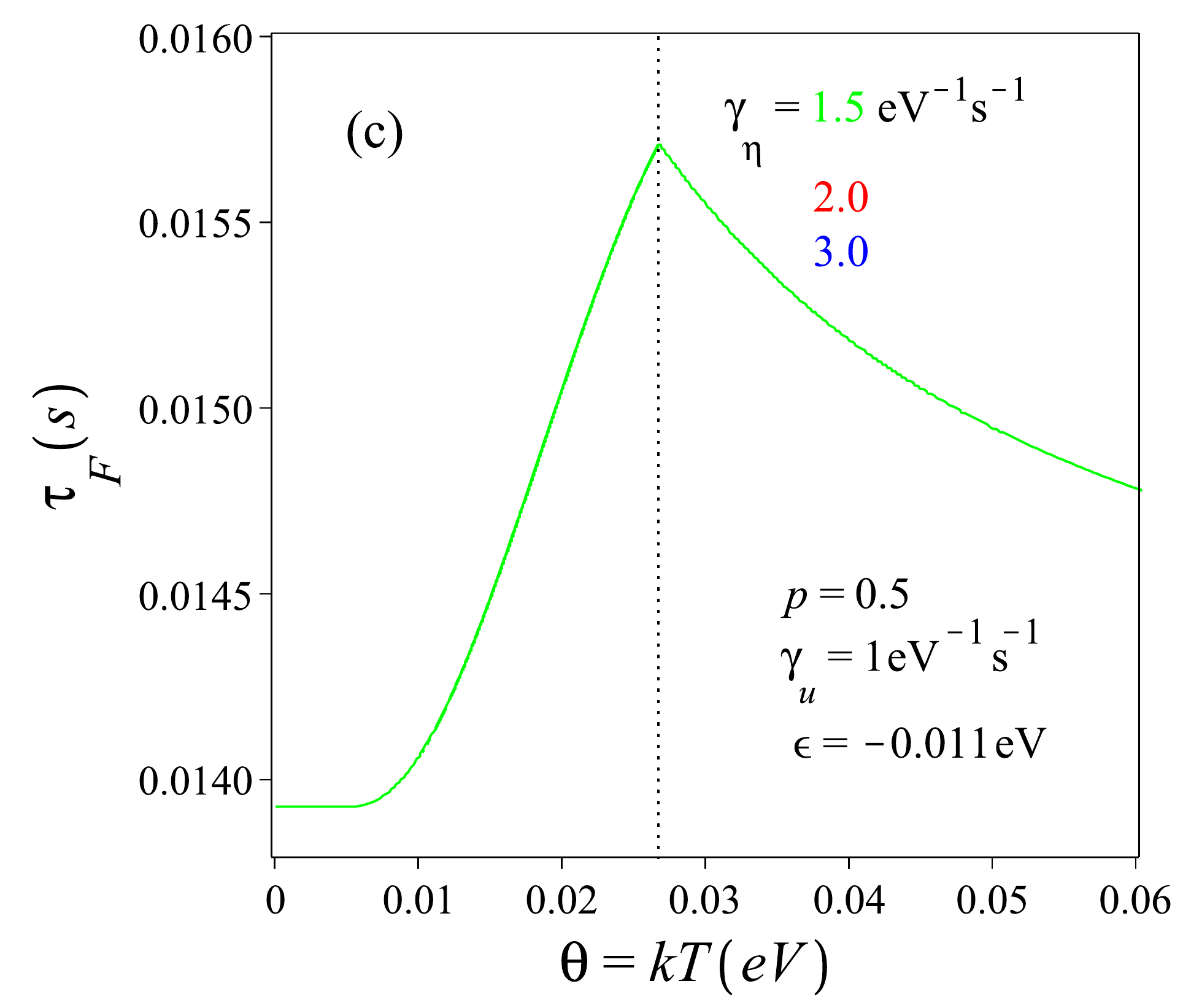}}\\
		%			\subfloat
		{\includegraphics[width=0.35\textwidth]{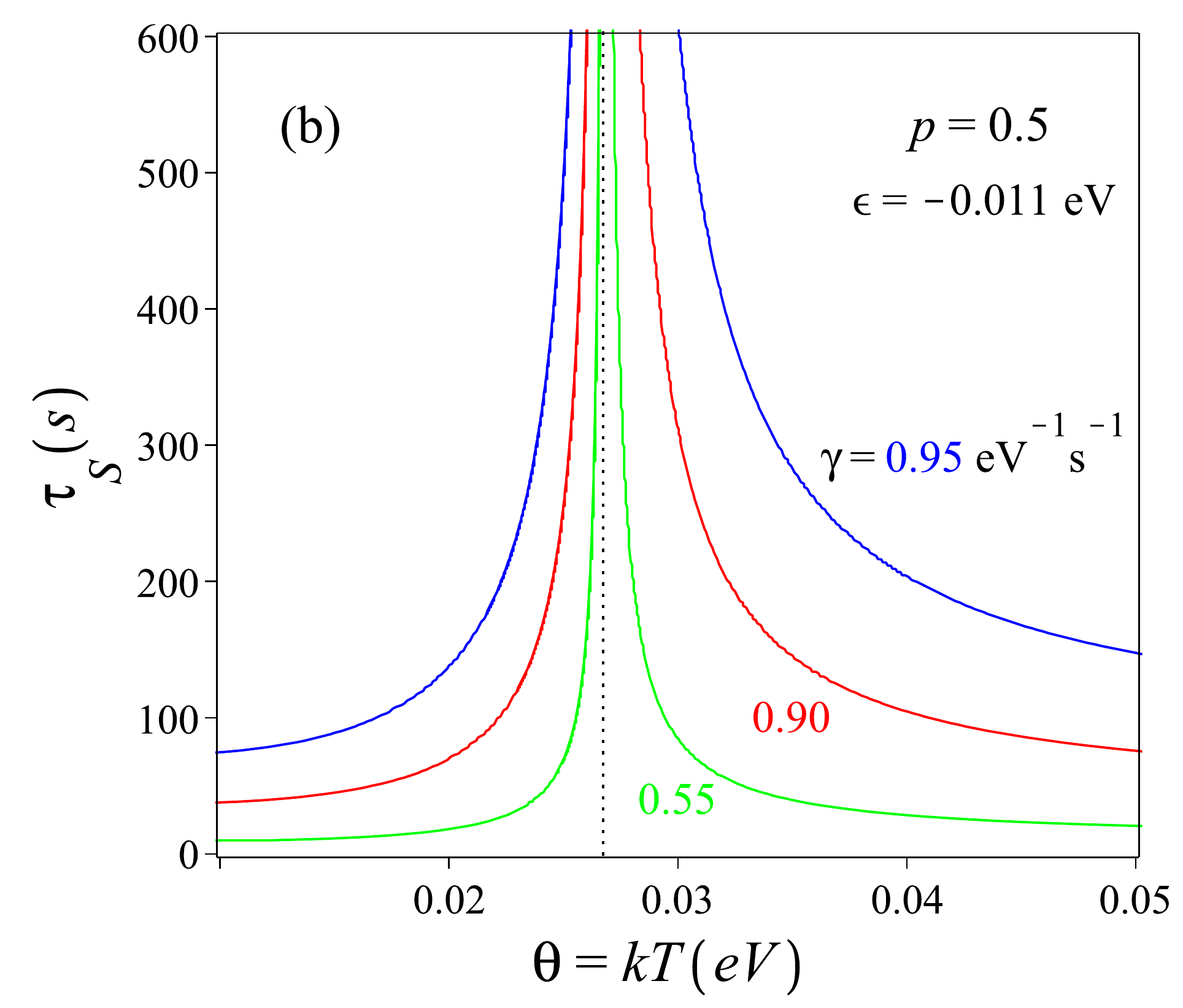}}
		%			\subfloat
		{\includegraphics[width=0.35\textwidth]{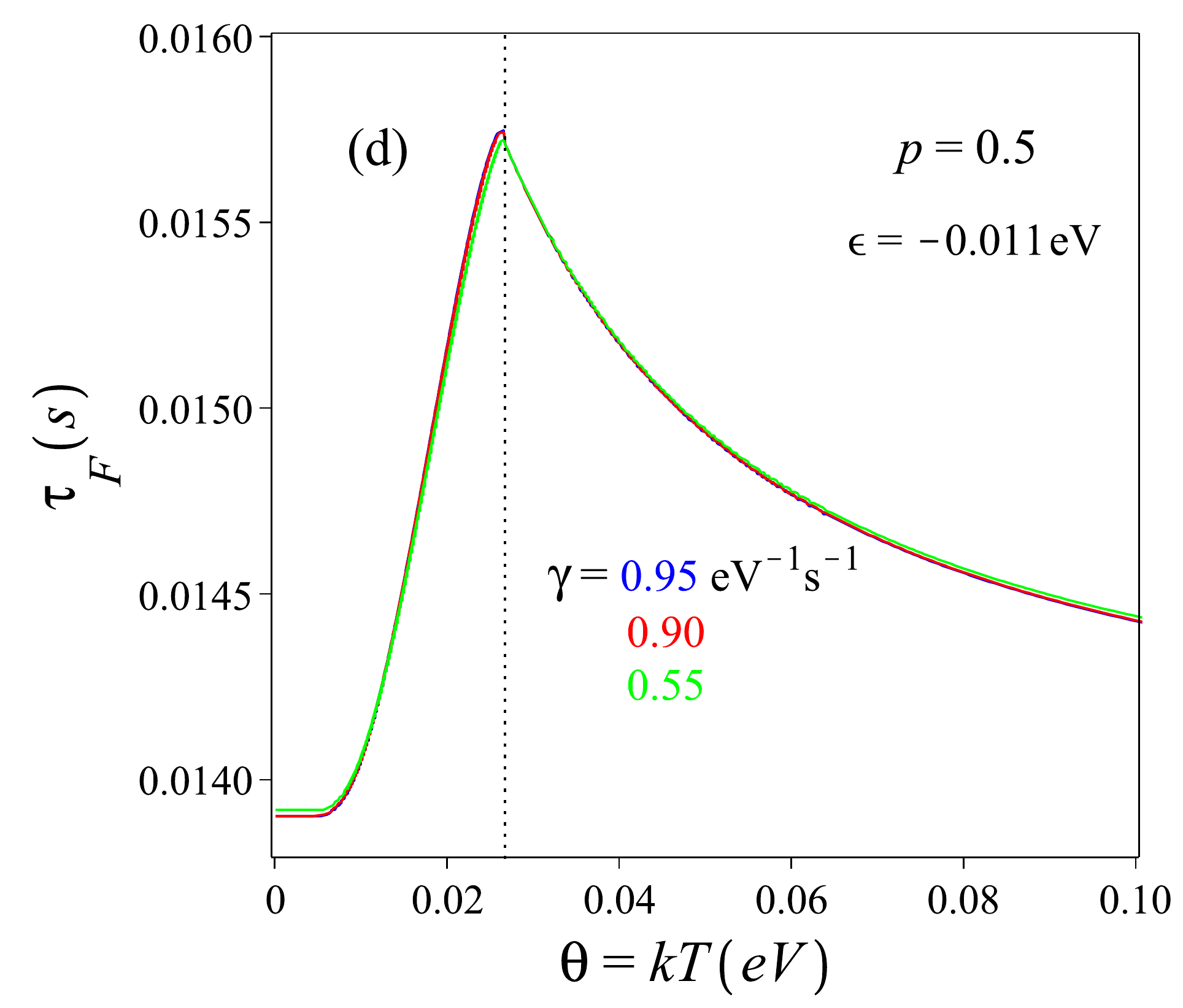}}
		\caption{(Colour online) (a), (b) Slow relaxation time $\tau_{S}$ and (c), (d) fast relaxation time $\tau_{F}$ vs. temperature $\theta$ in the absence and in the presence of the off-diagonal Onsager coefficient for $\varepsilon = -0.011$~eV, $p = 0.5$ GPa.}
		\label{fig4}
	\end{figure}
					
		Finally, the dependence of $\tau_{S}$  and $\tau_{F}$ on the temperature $\theta$ can be followed in figure~\ref{fig4}. These quantities are plotted for both $\gamma = 0$ and $\gamma \neq 0$ characterizing the speed of the relaxation process when   $\varepsilon=-0.011$~eV, $p=0.5$~GPa. It is clearly visible that the increases of $\gamma_{\eta}(\gamma$) result in much faster (slower) increase of $\tau_{S}$  around the critical temperature during the critical slowing down process [figures~\ref{fig4}(a) and \ref{fig4}(b)]. These are in direct agreement with previous results gained from dynamic studies for similar physical systems \cite{31,32,33,34,35}. It is quite evident from figures~\ref{fig4}(c) and \ref{fig4}(d) that the investigated plots of $\tau_{F}$ display a cusp-singularity at the critical temperature regardless of the value of rate parameters. By comparison of the colored curves in figures~\ref{fig4}(c) and \ref{fig4}(d), it can be stated that $\tau_{F}$ is not sensitive to any change in $\gamma_{\eta}$ when $\gamma = 0$ and in $\gamma$ for $\gamma \ne 0$. 
		\newpage			
		\section{Conclusions}
		In this work, we have studied the near equilibrium relaxation dynamics of the QLM for the ferroelectric SPS crystals by means of ORT. More particularly, time derivatives of polarization $\eta$  and volume deformation $u$ are treated as fluxes or currents conjugate to their appropriate forces. The forces are found using the mean-field free energy production. A set of rate (or kinetic) equations are derived from the linear relation between the forces and currents. In terms of the phenomenological constants, the solutions of rate equations are expressed as a set of two relaxations times denoted by $\tau_{S}$  and $\tau_{F}$. Using the parameter values related to the SPS crystals in \cite{24,25}, we have plotted these two quantities as a function of the energy gap, external pressure and temperature. Although the present calculations on $\tau_{S}$  and $\tau_{F}$  display exactly the same critical and tricritical behaviors as in the earlier investigations, we observe unusual results during the discontinous phase transition. In other words, the slow relaxation time $\tau_{S}$  shows a large jump in the $\tau_{S}$ vs. $\varepsilon$  and  $\tau_{S}$  vs. $p$ plots while fast relaxation time $\tau_{F}$ has a cusp-singularity and small jump in the $\tau_{F}$ vs. $\varepsilon$ and $\tau_{F}$ vs. $p$ plots, respectively, at the first-order phase transition. In particular, comparing figures~\ref{fig1}(b) and \ref{fig1}(d) with corresponding plots in \cite{32} it can be stated that this qualitative difference distinguishes our results from the earlier ORT calculations.

%		
%		\section*{Declaration of competing interest}
%		The authors declare that they have no known competing financial interests or personal relationships that could have appeared to influence the work reported in this paper.
%				

	\ukrainianpart
	
	\title
	{Часи повільної та швидкої релаксації квантової ґраткової моделі з локальними багатомінімумними потенціалами: феноменологічна динаміка в сегнетоелектричних кристалах  Sn$_{2}$P$_{2}$S$_{6}$%
	}
	\author[Р. Ердем, С. Озюм, Н. Гюджлю]{Р. Ердем\refaddr{label1},
		С. Озюм\refaddr{label2}, Н. Гюджлюр\refaddr{label3}}
	\addresses{
		\addr{label1} Фізичний факультет, Університет Акденіз, 07058, Анталія, Туреччина
		\addr{label2} Професійна школа Аласа Авні Джелік, Університет Гітіт, 19600, Чорум, Туреччина
		\addr{label3} Факультет фізичної освіти, Університет Неджметтина Ербакана, 42090, Конья, Туреччина
	}

	\makeukrtitle
	
	\begin{abstract}
		У продовження опублікованої раніше роботи [Velychko O. V., Stasyuk I. V., Phase Transitions, 2019, \textbf{92}, 420], для деформованої Sn$_{2}$P$_{2}$S$_{6}$ сегнетоелектричної ґратки наведено феноменологічну схему для розгляду ре\-лак\-са\-цiй\-ної динаміки квантової ґраткової моделі з багатомінімумними потенціалами. Схема базується на поєднанні статистичної рівноважної теорії та необоротної термодинаміки. З метою отримання узгодженого опису припускається, що дипольне впорядкування чи поляризацію  ($\eta$) та об'ємну деформацію ($u$) можна розглядати як потоки та сили в розумінні теорії Онзагера. З лінійних співвідношень між силами та потоками отримано рівняння динаміки, які характеризуються двома часами релаксації ($\tau_{S}, \tau_{F}$), що описують необоротній процес між рівноважними станами. Вивчено поведінку $\tau_{S}$ і $\tau_{F}$ поблизу сегнетоелектричних фазових переходів.  
		\keywords квантова ґраткова модель, сегнетоелектричні кристали, Sn$_{2}$P$_{2}$S$_{6}$, часи релаксації, теорія Онзагера
	\end{abstract}

	\end{document}